\documentclass[prl,superscriptaddress,twocolumn]{revtex4}%
\usepackage{amsfonts}
\usepackage{amsmath}
\usepackage{amssymb}
\usepackage[dvips]{graphicx}
\usepackage[dvips]{color}
\usepackage{epsfig}
\usepackage{dsfont}%

\newcommand{\nbar} {\overline{n}}

\newcommand{\ket}[1]{ {\left| #1 \right\rangle} }

\begin{document}

\title{Collapse and Revival of `Schr\"{o}dinger Cat' States}

\author{C. E. A. Jarvis}
\email{catherine.jarvis@bristol.ac.uk}
\affiliation{H H Wills Physics Laboratory, University of Bristol, Bristol BS8 1TL, United Kingdom}
\author{D. A. Rodrigues}
\affiliation{School of Physics and Astronomy, University of Nottingham, Nottingham, NG7
2RD, United Kingdom}
\author{B. L. Gy\"{o}rffy}
\affiliation{H H Wills Physics Laboratory, University of Bristol, Bristol BS8 1TL, United Kingdom}
\author{T. P. Spiller}
\affiliation{Hewlett Packard Laboratories, Filton Road, Bristol, BS34 8QZ, United Kingdom}
\author{A. J. Short}
\affiliation{DAMPT, Center of Mathematical Sciences, Wilberforce Road, Cambridge, CB3 0WA,
United Kingdom}
\author{J. F. Annett}
\affiliation{H H Wills Physics Laboratory, University of Bristol, Bristol BS8 1TL, United Kingdom}

\begin{abstract}
We study the dynamics of the Jaynes-Cummings Model for an array of $N_q$ two level
systems (or qubits) interacting with a quantized single mode electromagnetic
cavity (or quantum bus). For an initial
cavity coherent state $\left\vert \alpha \right\rangle$ and
the qubit system in a specified `basin of attraction' in its Hilbert space,
we demonstrate the oscillation of a superposition of two macroscopic quantum
states between the qubit system and the field mode. From the perspective of
either the qubit or the field system, there is collapse and revival of a
`Schr\"{o}dinger Cat' state.
\end{abstract}

\maketitle

Quantum superposition of macroscopically different states of matter are of
fundamental conceptual interest in many fields of physics such as
Measurement Theory \cite{Wheeler}, Quantum Optics \cite{GerryKnight}, Macroscopic Quantum
Tunnelling \cite{Takagi} and Quantum Computation \cite{NielsenChng}. Many of the relevant
experiments revolve around the preparation of and measurement on a specific
class of states called `Schr\"{o}dinger Cat' states \cite{Haroche}. These are
quantum superpositions of states which correspond to two (or more)
different values of a macroscopic variable, such as the magnetization or the
electric field in the cases of a large spin cluster, or large photon number or
phase in a cavity mode. In this letter we report the
surprising discovery that in a Jaynes-Cummings model (JCM), which describes an
array of two-level systems (qubits) interacting with a single cavity mode,
the time evolution can be such that sometimes the radiation field and sometimes
the qubit subsystem is in a `Schr\"{o}dinger Cat' state. In other words, a
superposition of macroscopically different states can shift from
one set of physical variables to another, as a function of time.

The quantum dynamics of two-level systems (qubits), coupled to a single
mode of an electromagnetic cavity, arise in many different physically interesting systems.
These include Rydberg atoms \cite{Haroche}, NMR studies of atomic nuclei \cite{Slichter,Pryadko}, Cooper Pair Boxes \cite{Wallraff}, Cavity Quantum Electrodynamics
\cite{Berman}, trapped ions \cite{Cirac}
and Quantum Computing \cite{NielsenChng}.  A very
general and simple Hamiltonian that captures the
relevant physics in all these fields is the JCM \cite{JaynesCum} (for one qubit)
and its generalization for multi-qubit
systems by Tavis and Cummings \cite{TavisCum}. Thus, our results are
pertinent to a broad range of physical systems.

One of the most interesting and surprising predictions of the JCM is the
`collapse and revival' of Rabi oscillations of the occupation probabilities
for various qubit states as the system evolves, from an initial state which
is a product of a coherent state $\left\vert \alpha \right\rangle $ for the
radiation field, and a generic qubit state $\left\vert \psi
_{N_{q}}\right\rangle $ \cite{GerryKnight}, where $N_q$ is the number of qubits. These remarkable dynamics occur
only because both the matter
and the cavity field are treated fully quantum mechanically. Indeed, our aim
here is to study the `collapse and revival' of `Schr\"{o}dinger cat'-like
states in a multi-qubit subsystem, as well as in the cavity field .

For clarity let us specify the multi-qubit JCM Hamiltonian, where each
qubit labelled $i$ has ground (excited) state ${|g_{i}\rangle }$ (${|e_{i}\rangle }$)
with energy $\epsilon _{g,i}$ ($\epsilon _{e,i}$). Up to a constant, the Hamiltonian
has the form%
\begin{eqnarray}
\hat{H} &=&\hbar \omega \hat{a}^{\dag }\hat{a}+\frac{\hbar }{2}%
\sum_{i=1}^{N_{q}}\Omega _{i}\hat{\sigma}_{i}^{z}+\hbar
\sum_{i=1}^{N_{q}}\lambda _{i}\left( \hat{a}\hat{\sigma}_{i}^{+}+\hat{a}%
^{\dag }\hat{\sigma}_{i}^{-}\right)  \label{eq:Nqham}
\end{eqnarray}%
where
$\hat{\sigma}_{i}^{z} ={\left\vert e_{i}\right\rangle }{\left\langle
e_{i}\right\vert }-{\left\vert g_{i}\right\rangle }{\left\langle
g_{i}\right\vert }$, $\hat{\sigma}_{i}^{+}={\left\vert e_{i}\right\rangle }{%
\left\langle g_{i}\right\vert }$, $\hat{\sigma}_{i}^{-}={\left\vert
g_{i}\right\rangle }{\left\langle e_{i}\right\vert }$
and $\hat{a}^{\dag }$ ($\hat{a}$) is the creation (annihilation)
operator for a photon with frequency $\omega $. The cavity-qubit$_{i}$
coupling constant is $\lambda _{i}$ and $\hbar \Omega _{i}=\epsilon
_{e,i}-\epsilon _{g,i}$. Here we consider only the cases of resonance, so
$\omega =\Omega _{i}$ for all $i$, and uniform
coupling, so $\lambda _{i}=\lambda $.

\begin{figure}[ptb]
\centering
{\epsfig{file=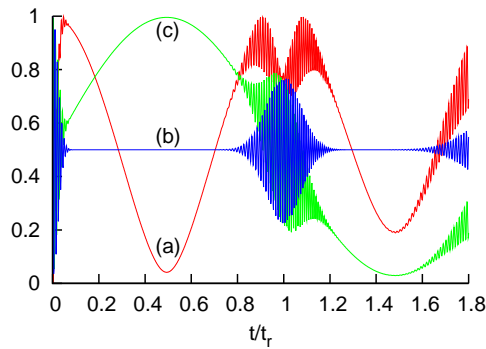, height=5cm} }\caption{(color online) Time evolution for a system with one qubit. (a) the entropy
of the qubit. (b) the probability of being
in the qubit's initial state ${\left|  g \right\rangle } $.  (c) the probability of being in the state ${\left|
\psi\right\rangle } _{att}^{+}$. At $t_{r}/2$ the probability of
being in the `attractor' state goes to one while the entropy goes to
zero. The qubit starts in the initial state $\ket{g}$ and the value of $\bar{n}=50$. }%
\label{fig:oneqexp}%
\end{figure}

The celebrated `collapse and revival' can be observed in the one-qubit case \cite{GerryKnight}.
It follows from an initial system state of
${\left\vert \Psi _{1}(0)\right\rangle }={\left\vert \psi _{1}\right\rangle }{%
\left\vert \alpha \right\rangle }$,
where ${\left\vert \alpha \right\rangle }=e^{-{\left\vert \alpha \right\vert
}^{2}/2}\sum_{n=0}^{\infty }\frac{\alpha ^{n}}{\sqrt{n!}}{\left\vert
n\right\rangle }$, $\alpha =\sqrt{\overline{n}}e^{-i\theta }$ , $\overline{n}$ is
the average number of photons in the field and
${|\psi_{1}\rangle }=(C_{g}{|g\rangle }+C_{e}{|e\rangle })$. The Rabi oscillations (in the
probability of the qubit being in its initial state) demonstrate collapse, on a
time scale of $t_{c}\simeq \frac{\sqrt{2}}{\lambda }$, and then revival at
$t_{r}\simeq \frac{2\pi \sqrt{\overline{n}}}{\lambda }$. This is illustrated
in Fig.~\ref{fig:oneqexp} for $C_{g}=1$, $C_{e}=0$ by plotting
$\sum_{n=0}^{\infty }{\left\vert {\left\langle g,n|\Psi _{1}(t)\right\rangle }%
\right\vert }^{2}$, where $\left\langle g,n\right\vert $ corresponds to the
qubit in its ground state with $n$ photons in the cavity.

A second notable feature of this time evolution, discovered by
Gea-Banacloche \cite{Ban90}, is that at $t=\frac{1}{2}t_{r}$, $\left\vert \Psi
_{1}(\frac{1}{2}t_{r})\right\rangle $ again factorises into a qubit part ${%
|\psi \rangle }_{att}^{+}$ and a cavity part ${|\Phi (\frac{1}{2}%
t_{r})\rangle }$. Moreover, remarkably, the former is given by%
\begin{equation}
{|\psi \rangle }_{att}^{\pm }=\frac{1}{\sqrt{2}}\left( e^{-i\theta }{%
|e\rangle }\pm i{|g\rangle }\right)  \label{eq:1qattractor}
\end{equation}%
where $\theta $ is the phase of the initial coherent state, for \emph{all}
initial conditions such that $\left\vert C_{g}\right\vert ^{2}+\left\vert
C_{e}\right\vert ^{2}=1$. Under same conditions
${|\psi\rangle }_{att}^{-}$ is attained at $t=\frac{3}{2}t_{r}$. Because of this
strikingly non-linear behavior, following Phoenix and Knight \cite{PheKnight91},
we shall refer to the states Eq. (\ref{eq:1qattractor})
as `attractors'. The probability that the qubit is in the state ${|\psi
\rangle }_{att}^{+}$, given by $\sum_{n=0}^{\infty }\left\vert
\left\langle \psi _{att}^{+},n|\Psi _{1}(t)\right\rangle \right\vert
^{2}$ is also shown in Fig.~\ref{fig:oneqexp}, together with the von Neumann
entropy $S^{q}(t)=-\text{Tr}\left( \rho ^{q}(t)\ln \rho ^{q}(t)\right)$
associated with the qubit density matrix $\rho ^{q}(t)=\text{Tr}_{F}\left(
{\left\vert \Psi _{1}(t)\right\rangle }{\left\langle \Psi _{1}(t)\right\vert
}\right)$, reduced by tracing over the field. Clearly, at $t={\frac{1}{2}t_{r}}$
the entropy $S^{q}(t)$ approaches zero, and the analytical solution shows that in the limit $\overline{n}
\rightarrow \infty $ the entropy goes to zero, indicating that the radiation field and the qubit are
not entangled \cite{Ban90}.

Prompted by these results, we have investigated the $N_{q}>1$ qubit
evolution, starting in a state ${\left\vert \Psi _{N_{q}}(0)\right\rangle }
={\left\vert \psi _{N_{q}}\right\rangle }{\left\vert \alpha \right\rangle }$.
Analytically, in the large $\nbar$ limit, we have found that the states
\begin{equation}\label{eq:nqattract}
{|\psi _{N_{q}}\rangle }_{att}^{\pm }=\frac{1}{\sqrt{2^{N_{q}}}}\left(
e^{-i\theta }{|e\rangle }\pm i{|g}\rangle \right) ^{\otimes N_{q}}
\end{equation}%
can also be regarded as `attractors' in a similar, dynamical, sense as
outlined above. The only difference is that in the $N_{q}>1$ case,
${|\psi_{N_{q}}\rangle }_{att}^{\pm }$ only occurs at $t=t_{r}/2N_{q}$
for a restricted range of initial conditions, which we shall term the `basin
of attraction'. As before, at half way to revival of the initial state,
namely $t=t_{r}/2N_{q}$, the radiation field and the qubit system are not
entangled. However, now the question of entanglement between the qubits
arises. Compared to the one-qubit problem studied by Gea-Banacloche \cite{Ban91} this
is a new feature of the multi-qubit case. Clearly, as
${|\psi_{N_{q}}\rangle }_{att}^{\pm }$ is a simple product of individual
qubit states when the qubit subsystem is in this state, the qubits are not
entangled with each other. This is surprising because, as we shall show later,
this attractor state can be reached from initial states with arbitrary
entanglement between qubits.

Before examining how such interesting dynamics can occur, it is useful to
note that the above product state is a spin coherent state for a finite
$N_{q}$-qubit system. Following Radcliffe \cite{Radcliffe} we define such states as
\begin{equation}\label{eq:qcoherent}
\left\vert \beta ,N_{q}\right\rangle=\frac{1}{{\cal N}}
\sum_{m=N_q/2}^{N_q/2}\sqrt{{\cal C}_{\frac{N_q}{2}+m}^{N_q}}
\beta^{\frac{N_q}{2}-m}\ket{N_{q},m}
\end{equation}
where the states $\left\vert N_{q},m\right\rangle $ are the fully
symmetrized $N_{q}$ qubit states, for $N_{e}$ qubits excited
and $N_{g}$ in the ground state, with $m=\frac{N_{e}-N_{g}}{2}$ and $\beta $ a
complex number that characterises the the state.
The normalization and combinatoric factors are
${\cal N}=\left(1+\left\vert \beta\right\vert^2 \right)^{N_{q}/2}$ and
${\cal C}_{\frac{N_q}{2}+m}^{N_q}=\frac{N_q!}{\left(\frac{N_q}{2}-m\right)!\left(\frac{N_q}{2}+m\right)!}$.
As can be readily shown, the
`attractor' states identified in Eq. \ref{eq:nqattract} are given by
${\left\vert \beta =\pm ie^{i\theta },N_{q}\right\rangle }$.
Below we explore the implications of this observation for the dynamics of
qubit states in the `basin of attraction'.

\begin{figure}[tbp]
\centering
\epsfig{file=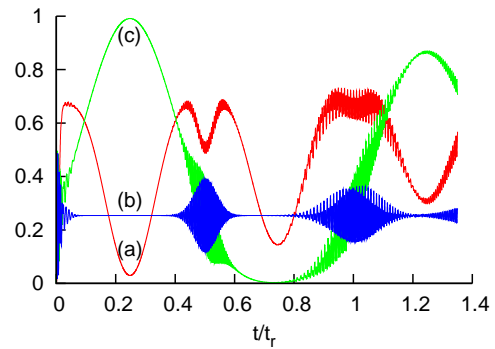,height=5cm}
\caption{(color online) Time evolution for a system with two qubits. (a) the
entropy of the qubits. (b) the probability of the two qubit state ${%
\left\vert gg\right\rangle }$. (c) the probability of being in the two qubit
`attractor' state ${\left\vert \protect\psi _{2}\right\rangle }%
_{att}^{+}$ when the initial phase of the radiation field is $\protect%
\theta =0$. The two qubit `attractor' state is reached at $t_{r}/4$. The
initial state of the qubits is $\frac{1}{\protect\sqrt{2}}({\left\vert
ee\right\rangle }+{\left\vert gg\right\rangle })$ and the value of $%
\overline{n}=50$.}
\label{fig:twoqubitatt}
\end{figure}

Let us now investigate the `basin of attraction' for the simplest
multi-qubit system, with $N_{q}=2$. In this case the time evolution described
by $\left\vert \Psi _{2}(t)\right\rangle $ is readily found \cite{Chumakov95}.
For the most general, normalized, initial state
\begin{equation}
\left\vert \psi _{2}\right\rangle =C_{ee}\left\vert ee\right\rangle
+C_{eg}\left\vert eg\right\rangle +C_{ge}\left\vert ge\right\rangle
+C_{gg}\left\vert gg\right\rangle
\end{equation}
the exact analytical solution will be given elsewhere \cite{Jarvis08b}. Here
we consider only the sector determined by the restrictions: $a=e^{i\theta
}C_{ee}=e^{-i\theta }C_{gg}$ and $\sqrt{\frac{1}{2}-{\left\vert a\right\vert
}^{2}}=C_{eg}=C_{ge}$. As will be illustrated presently, these define the
`basin of attraction' for the `attractor' ${|\psi _{2}\rangle }%
_{att}^{+}$. Namely, for any complex number $a$ satisfying the
condition $0\leq \left\vert a\right\vert \leq 1/\sqrt{2}$ in
\begin{equation}
{\left\vert \psi _{2}\right\rangle }=a\left( e^{-i\theta }{\left\vert
ee\right\rangle }+e^{i\theta }{\left\vert gg\right\rangle }\right) +\sqrt{%
\frac{1}{2}-{\left\vert a\right\vert }^{2}}\left( {\left\vert
eg\right\rangle }+{\left\vert ge\right\rangle }\right),
\label{eq:initattract}
\end{equation}%
the probability (given by $P_{2\,att}(t)=\left\langle \psi
_{2\,att}^{+}\right\vert \rho ^{q}(t)|\psi _{2\,att}^{+}\rangle $%
) that the two qubits are in the state ${|\psi _{2}\rangle }%
_{att}^{+}$ approaches unity at some time $t^{\ast }$. An example of such behavior (for
$\theta =0$) is shown in Fig.~\ref{fig:twoqubitatt}. To highlight the
similarity with the analogous phenomenon in the one qubit case
(Fig.~\ref{fig:oneqexp}), we also show the entropy $S^{q}(t)$. This is calculated from
$\rho^{q}$, the two-qubit density matrix reduced with respect to the cavity
field coordinate, which describes a mixed state for most times $t$. Notably,
at $t^{\ast }=\frac{1}{4}t_{r}$, where $P_{2\,att}(t)=1,$ the entropy $%
S^{q}(t)$ approaches zero in the large $\overline{n}$ limit, indicating that
the system of two qubits is not entangled with the field.

\begin{figure}[ptb]
\centering{\epsfig{file=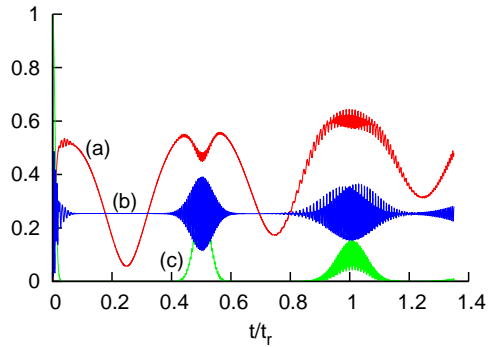,height=5cm} }\caption{(color online) The qubit system
started in the maximally entangled state $({\left\vert ee\right\rangle
}+{\left\vert gg\right\rangle })/\sqrt{2}$ and $\nbar=50$. (a) the entropy of the qubit system. (b) the
probability of being in the state ${\left\vert gg\right\rangle }$. (c) the mixed state tangle of the qubit
system.}%
\label{fig:reventanglement}%
\end{figure}

As we have already noted, the interesting new feature of the two-qubit case
as opposed to the one-qubit case is that the former is in general host to
entanglement between qubits and this provides an opportunity to study the
dynamics of such entanglement. For example, whilst almost all of the initial
states in the `basin of attraction' given in Eq. (\ref{eq:initattract})
describe entangled qubits, they all evolve into ${|\psi _{2}\rangle }%
_{att}^{+}$ at $t=\frac{1}{4}t_{r}$ where they are not entangled. We
investigated the pure state tangle of the initial condition defined as $\tau
=4\left\vert C_{ee}C_{gg}-C_{eg}C_{ge}\right\vert ^{2}$ \cite{Wooters98} as
a function of $a$ and found that although there are
only two points where $\tau =0$, all values of entanglement, including $\tau
=1$ indicating maximal entanglement, are present in the `basin of attraction'.
Thus we are observing the time evolution of a generic amount of
entanglement. To throw further light on the matter, in
Fig. \ref{fig:reventanglement} we show the time evolution of the mixed state tangle \cite{Munro}
calculated from $\rho ^{q}$ for the maximally entangled initial state $|\psi
_{2}\rangle =\frac{1}{\sqrt{2}}\left( \left\vert ee\right\rangle +\left\vert
gg\right\rangle \right) $.
Evidently, just as the occupation of the initial qubit states collapses and
revives, so does the entanglement. This phenomenon was first noted by
Rodrigues \emph{et al.} in a similar context \cite{Rodrigues}.

Interestingly, not only the `attractor states', ${|\psi _{N_{q}}\rangle }%
_{att}^{\pm }$, but also their `basin of attraction' can be found for
the general $N_{q}$-qubit case. In fact, using the large $\overline{n}$
expansion of Meunier \emph{et al.} \cite{Meunier} we found the `basin of attraction' for all
the attractor states identified in Eq. \ref{eq:nqattract}. We have established that, for
a given $N_{q}$, only initial states parametrized as follows:
\begin{eqnarray}
{\left\vert \psi _{N_{q}}\right\rangle }_{a} &=&\sum_{m=-N_q/2}^{N_{q}/2}\frac{%
A(N_{q},a)e^{-i(\frac{N_{q}}{2}-m)\theta }\sqrt{N_{q}!}}{\sqrt{\left(\frac{N_q}{2}+m\right)!\left(\frac{N_q}{2}-m\right)!}}%
{\left\vert N_{q},m\right\rangle }  \nonumber
\label{eq:Nqbasin} \\
A(N_{q},a) &=&\left\{
\begin{array}{cc}
a & \text{if $k$ is even} \\
&  \\
\sqrt{\frac{1}{2^{N_{q}-1}}-{\left\vert a\right\vert }^{2}} & \text{if $k$
is odd}%
\end{array}%
\right.
\end{eqnarray}%
where $0\leq {\left\vert a\right\vert }\leq \frac{1}{\sqrt{2^{N_{q}-1}}}$ ,
will evolve into the spin coherent state ${\left\vert \beta =ie^{-i\theta
},N_{q}\right\rangle }$ at the time $t=t_{r}/2N_{q}$.

A remarkable feature of this `basin of attraction' is that it includes
`Schr\"{o}dinger cat' states of the qubit system. Indeed it can be readily shown,
by carrying out the sum over $m$ in Eq. (\ref{eq:Nqbasin}) and
identifying two different spin coherent states, that
\begin{eqnarray}
&&{\left\vert \psi _{N_{q}}\right\rangle }_{a} =\sqrt{2^{Nq-2}}\left( a+%
\sqrt{\frac{1}{2^{N_{q}-1}}-{\ {\left\vert a\right\vert }}^{2}}\right) {%
\left\vert e^{-i\theta },N_{q}\right\rangle } \nonumber\\
&&\ \ \ \ \ \ \ \ +\sqrt{2^{Nq-2}}\left( a-\sqrt{\frac{1}{2^{N_{q}-1}}-{\ {\left\vert
a\right\vert }}^{2}}\right) {\left\vert-e^{-i\theta
},N_{q}\right\rangle }\nonumber\\
\end{eqnarray}
Therefore the `basin of attraction' consists of linear superpositions of
two quite different spin-coherent states, Eq. \ref{eq:qcoherent}, ${\left\vert e^{-i\theta
},N_{q}\right\rangle }$  and${\left\vert -e^{-i\theta }N_{q},\right\rangle}$.
For $N_{q}\leq 3$ these two states are hard to distinguish,
but for large $N_{q}$ they can be regarded as macroscopically
different and hence their
superposition is very similar to the finite $N_{q}$-qubit `Schr\"{o}dinger
cat' state with equal coefficients. Indeed for $a=0$ and $a=\frac{1}{\sqrt{2^{N_q-1}}}$,
${\left\vert \psi _{N_{q}}\right\rangle }%
_{a=0}\ =\frac{1}{\sqrt{2}}\ \left( \left\vert \beta ,N_q\right\rangle
+\left\vert -\beta ,N_q\right\rangle \right) \equiv \left\vert \beta,N_q
\right\rangle _{Sch}$ for $\beta =e^{-i\theta }$.

\begin{figure}[ptb]
\centering{\epsfig{file=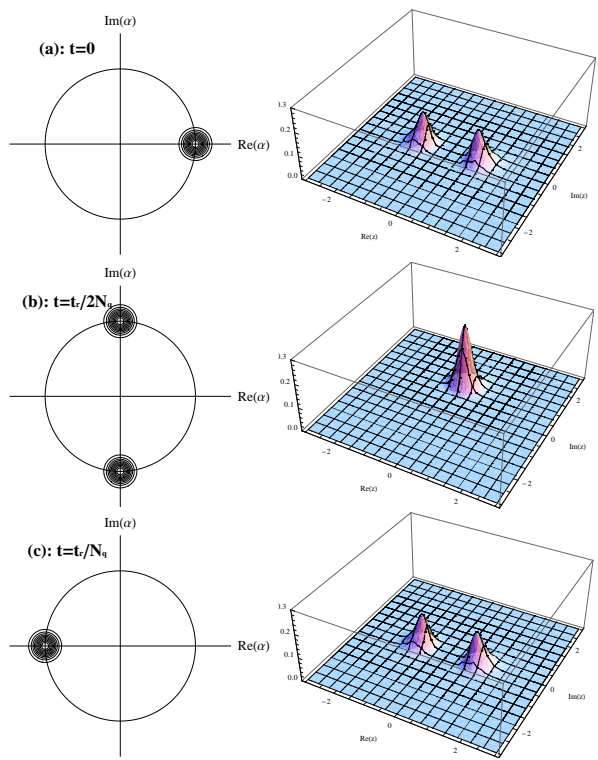,height=10cm} }\caption{(color online)
Diagrams of the $Q$ function \cite{GardinerZoller} (left) and spin $Q$ function \cite{spinqfunc} (right) at three
different times. (a) the time $t=0$, where the cavity is in a coherent state
and the qubits are in a spin Schr\"{o}dinger cat state. (b) the time $t=t_r/2N_q$
which is the time of the first attractor. (c) the time $t=t_r/N_q$, when the field
states are again overlapping and in a coherent state. They show the `Schr\"{o}dinger Cat' state moving from the qubits to the radiation field and back again in the limit
$\nbar\rightarrow\infty$, $N_q=40$, $\theta=0$. The $Q$ function for the field
has been scaled to a unit circle.}%
\label{fig:catmove}%
\end{figure}

Thus, the dynamics governed by the JCM Hamiltonian can transfer
an initial state which is a product of a coherent state for the radiation
field, ${\left\vert \alpha \right\rangle }$, and a highly quantum mechanical
`Schr\"{o}dinger cat'-like state for the qubits, at the time $t^*=\frac{1}{%
2N_{q}}$, into another product state where the qubit component is a rather
classical qubit-(spin) coherent state, ${\left\vert \beta \right\rangle }$,
with no entanglement between the qubits. Given this intriguing time
evolution, it is interesting to examine the state of the radiation field
at $t^*$. Again in the limit of large average photon number $\overline{n}$ in
the cavity we find%
\begin{eqnarray}
&&{\left\vert \Phi _{N_{q}}\left(\frac{t_{r}}{2N_{q}}\right)\right\rangle } =\left[%
\left( a-\sqrt{\frac{1}{2^{N_{q}-1}}-{\ {\left\vert a\right\vert }}^{2}}%
\right) e^{i\pi \overline{n}/2}{\left\vert i\alpha \right\rangle }\right.\nonumber \\
&&\left. -\left( a+\sqrt{\frac{1}{2^{N_{q}-1}}-{\ {\left\vert
a\right\vert }}^{2}}\right) e^{-i\pi \overline{n}/2}{\left\vert -i\alpha
\right\rangle}\right]\sqrt{2^{N_{q}-2}}
\end{eqnarray}%
Evidently, this is a `Schr\"{o}dinger cat'-like state for the radiation
field. Hence, we may conclude that the initial quantum information which was
encoded in the qubit `Schr\"{o}dinger cat' state moves to and resides in
the radiation field at $t^*$. As might now be expected, the time evolution after $t^*$
returns this information to the qubits, as the initial state
revives. This remarkable time evolution is illustrated in Fig. \ref{fig:catmove}.

In summary, we have studied the dynamics governed by the JCM for an array of qubits
interacting with a single mode of quantized
radiation field, a quantum bus. We have shown that a product state, comprising a
coherent state for the photons and a macroscopic `Schr\"{o}dinger cat'-like state for
the qubits, transforms into a product of a `Schr\"{o}dinger cat'-like state
for the photons and a spin-coherent state for the qubits. Furthermore, as
time goes on this transformation is reversed and then the process starts all
over again. This suggests that some universal quantum information is being
passed back and forth between the two subsystems, each of which
manifests it through its own physics. Further study of
this intriguing `collapse and revival' phenomenon should include
an investigation of the effects
of decoherence and dissipation, of which initial studies have been made by Meunier \emph{et al.} \cite{Meunier}, and design of experiments to observe it.

\smallskip
\acknowledgments

The work of C.E.A.J. was supported by UK HP/EPSRC case studentship,
and D.A.R. was supported by EPSRC-GB grant no EP/D066417/1. A.J.S. acknowledges support from a Royal Society University Research Fellowship and the EC QAP project. We thank the ESF network AQDJJ for partial support.

\bibliography{MacroscopicCollapseandRevival}

\end{document}